\documentclass[%
reprint,
superscriptaddress,
amsmath,amssymb,
aps,
pre,
]{revtex4-2}

\usepackage{dcolumn}
\usepackage{natbib}
\usepackage{subcaption}
\usepackage{graphicx}
\usepackage{bm}
\usepackage{afterpage}
\usepackage{tabularx}
\usepackage{ragged2e}
\usepackage{xcolor}

\newcommand{\be}{\begin{equation}}
\newcommand{\ee}{\end{equation}}
\newcommand{\beqn}{\begin{eqnarray}}
\newcommand{\eeqn}{\end{eqnarray}}

\begin{document}

\title{Multicritical Infection Spreading}

\author{Leone~V.~Luzzatto}
\affiliation{Department of Physics and Astronomy, Northwestern University, Evanston, IL 60208}
\affiliation{NSF-Simons National Institute for Theory and Mathematics in Biology, Chicago, IL 60611}

\author{Juan~Felipe~Barrera~L\'opez}
\affiliation{Department of Physics and Astronomy, Northwestern University, Evanston, IL 60208}

\author{Istv\'an~A.~Kov\'acs}
\email{istvan.kovacs@northwestern.edu}
\affiliation{Department of Physics and Astronomy, Northwestern University, Evanston, IL 60208}
\affiliation{NSF-Simons National Institute for Theory and Mathematics in Biology, Chicago, IL 60611}
\affiliation{Northwestern Institute on Complex Systems, Northwestern University, Evanston, IL 60208}
\affiliation{Department of Engineering Sciences and Applied Mathematics, Northwestern University, Evanston, IL 60208}


\begin{abstract}
The contact process is a simple infection spreading model showcasing an out-of-equilibrium phase transition between a macroscopically active and an inactive phase. 
Such absorbing state phase transitions are often sensitive to the presence of quenched disorder.
Traditionally, a phase transition in the disordered contact process is either triggered by dilution or by locally varying the infection rate. However, when both factors play an important role, a multicritical point emerges that remains poorly understood. Here, we study the multicritical contact process by large-scale Monte Carlo simulations in two and three dimensions. 
The multicritical behavior is found to be universal and exhibits ultra-slow, activated dynamical scaling, with exponents consistent with those predicted by the strong disorder renormalization group method. This finding indicates that the multicritical contact process belongs to the same universality class as the multicritical quantum Ising model, in both two and three dimensions.
\end{abstract}


\maketitle


\section{Introduction}
Even with recent events, our understanding of infection spreading remains limited.
The contact process (CP) \cite{harris1974, liggett1999stochastic} is a simple stochastic model in which agents have the ability to infect their neighbors and, once infected, to become healthy again.
On a lattice, the CP is equivalent to the susceptible-infected-susceptible (SIS) model, motivated by scenarios in which agents can be infected multiple times.
Most theoretical and computational studies focus on the simplified case in which all agents share the same infection and recovery rates---a far cry from real-life scenarios.

In the absence of heterogeneities, the CP exhibits an out-of-equilibrium absorbing state phase transition between a macroscopically active and a fully inactive phase that falls into the universality class of directed percolation (DP) \cite{MarroDickman1999, Hinrichsen2000, henkel2008non, RevModPhys.76.663}.
A perturbative argument, known as the Harris criterion \cite{Harris1974criterion}, predicts that a phase transition is sensitive to weak time-independent quenched disorder in $d$ dimensions whenever $d\nu<2$, where $\nu$ is the correlation length critical exponent \cite{PhysRevLett.57.2999}. As a result, in the presence of sufficiently strong disorder, a qualitatively new emergent behavior is expected, characterized by a new exponent $\nu'\geq2/d$ \cite{PhysRevLett.57.2999}. 
For the DP universality class, the Harris criterion implies that the critical behavior is modified even by weak quenched disorder below four dimensions~\cite{KIM2022126464}.

\begin{figure}[t!]
\begin{center}
\includegraphics[width=2.75in,angle=0]{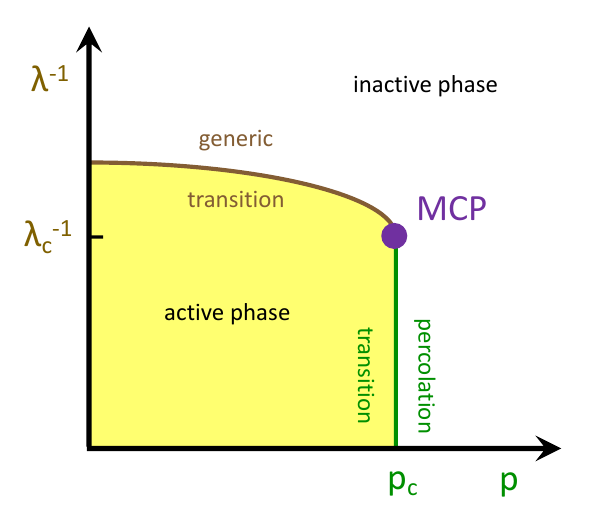}
\end{center}
\vskip -8mm
\caption{
\justifying \textbf{Schematic phase diagram of the disordered CP in two and three dimensions \cite{Vojta2009, Vojta2012}.}
The phase transition at the (bond) dilution $p=p_c$ (teal line) is controlled by the percolation fixed point, while below $p_c$ the generic disordered transition (brown) is observed by tuning the infection rate $\lambda$.
Both the percolation and generic universality classes are known examples of an IDFP (see text). In this paper, we characterize the distinct universality class at the multicritical point (MCP, purple), located at $p_c=1/2$, $\lambda_c=3.5585(20)$ for a two-dimensional square lattice and $p_c\approx0.751188$, $\lambda_c=5.600(5)$ for a three-dimensional cubic lattice.
At $p=0$, the model has a phase transition in the DP universality class, located at $\lambda_{c,0}=1.64877(3)$ and $\lambda_{c,0}=1.31686(1)$ in two and three dimensions, respectively \cite{Dickman1999}.
}
\label{fig:phase}
\end{figure}

Disorder is typically introduced to the CP either by randomly drawing individual infection (and/or healing) rates from a given distribution \cite{Noest1986, Hooyberghs2003} or by diluting the underlying lattice \cite{Moreira1996, Vojta2009}.
Depending on the introduced heterogeneities, two distinct non-equilibrium phase transitions can be observed in the disordered CP. A \emph{generic transition} takes place in the case of random infection rates or when the dilution of the lattice is below the percolation threshold; a \emph{percolation transition} is observed when the lattice is critically diluted.
In the phase diagram of the CP on a diluted lattice in dimensions $d\geq2$, shown in Fig.~\ref{fig:phase}, the two transitions meet at a multicritical point, which is the focus of this work.

\begin{table*}[t!]
\centering
\caption{\justifying\textbf{Critical exponents of the disordered CP.} Critical exponents at the generic disordered transition, percolation transition, and multicritical point (MCP). At the generic transition, two estimates of the exponents are shown, obtained through direct Monte Carlo (MC) simulations of the CP~\cite{Vojta2009,Vojta2012} and through numerical implementations of the SDRG to the quantum Ising model~\cite{Kovacs2010-2d-sdrg,Kovacs2011-3d-sdrg}. The critical exponents at the percolation transition are determined by the standard percolation exponents~\cite{Vojta2006}, which are known exactly in two dimensions and have been determined numerically to high precision in three dimensions~\cite{stauffer_book}. Exponents that do not apply to a given transition are left blank; ``$-$'' is used for exponents of unknown value. The numbers in parentheses give an estimate of the uncertainty on the last digits. Values of $\gamma$ are calculated from the other exponents using Eq.~(\ref{eq:gamma})---since this calculation uses $\nu_\Delta$ (and not $\nu_p$), we only report $\gamma$ at the generic disordered transition and at the multicritical point. 
}\label{tab1}

\begin{ruledtabular}
\begin{tabular}{cccccccccc}     
    & \multicolumn{4}{c}{$d=2$} & & \multicolumn{4}{c}{$d=3$} \\ 
    & Generic, & Generic, & Percolation & MCP & & Generic, & Generic, & Percolation & MCP \\
    & MC~\cite{Vojta2009} & SDRG~\cite{Kovacs2010-2d-sdrg} & \cite{stauffer_book} & (this~work) & & MC~\cite{Vojta2012} & SDRG~\cite{Kovacs2011-3d-sdrg} & \cite{stauffer_book} & (this~work) \\

\noalign{\vskip 1pt}
\hline
\noalign{\vskip 3pt}
    
    $\beta$         & 1.15(5)   & 1.22(4)   & $5/36 \approx0.14$    & 1.10(5)     & & 1.87(7)   & 1.82(4)   & 0.417 & 1.62(20) \\
    $\nu_\Delta$    & 1.20(15)  & 1.24(2)   &                       & 1.43(15)    & & 0.98(6)   & 0.99(2)   &       & 1.2(3)   \\
    $\nu_p$         &           &           & $4/3\approx1.33$      & $-$         & &           &           & 0.875 & $-$      \\   
    
    $\bar{\Theta}$  & 0.15(3)   & 0.08(7)   &                       & 0.58(7)     & & -2.1(2)   & -1.5(1)   &       & 0.22(14) \\
    $\psi$          & 0.51(6)   & 0.58(2)   & $91/48\approx1.9$     & 0.698(8)   & & 0.38(3)   & 0.46(2)   & 2.523 & 0.86(4)  \\
    $\bar{\delta}$  & 1.9(2)    & 2.22(2)   & $5/91\approx0.05$     & 1.11(4)     & & 5.0(2)    & 4.0(2)    & 0.188 & 1.56(9)  \\
    $\gamma$        & 0.10(4)   & 0.06(5)   &                       & 0.57(14)    & & -0.80(18) & -0.69(9)  &       & 0.26(19) \\

\end{tabular}
\end{ruledtabular}
\end{table*}

At both transitions, the DP critical behavior of the clean CP is replaced by ultra-slow, activated scaling---manifestation of an infinitely disordered fixed point (IDFP) \cite{IgloiMonthus2005,Igloi2018}---where the role of time is played instead by its logarithm.
While the percolation IDFP is characterized by the critical exponents of standard percolation \cite{Sachdev_2011}, the generic IDFP has a new set of ``quantum percolation'' critical exponents that are known through the so-called strong disorder renormalization group (SDRG) method \cite{IgloiMonthus2005,Igloi2018}.
Numerical values of the critical exponents at the percolation and generic disordered transitions of the CP are reported in Table~\ref{tab1}.

The SDRG approach was first developed for disordered quantum systems \cite{Ma1976, Dasgupta1980} and was shown by Daniel Fisher to be asymptotically exact in the vicinity of the critical point in one dimension \cite{Fisher1999, Pandey2023}.
To illustrate how the SDRG procedure can be applied to the absorbing-state critical point of the disordered CP, we sketch here the theoretical arguments presented in greater detail in Refs.~\cite{Hooyberghs2003,Hooyberghs2004}.
The first step is to treat the disordered CP as a Markov process with generator $\mathcal{H}$.
The steady state of this process corresponds to the ground state of a quantum system with $\mathcal{H}$ as its Hamiltonian.
The SDRG equations for the disordered CP can then be inferred from $\mathcal{H}$, and the resulting decimation prescriptions (\textit{i.e.}, the RG flow) are essentially identical to those of the disordered quantum Ising model, suggesting that the critical behavior of the two models is governed by the same IDFP---in $d<4$ dimensions and given strong enough initial disorder.
A similar SDRG treatment can be applied to the Hamiltonian form of random Reggeon field theory~\cite{browerQuantumSpinModel1977,Hooyberghs2004}, which describes the critical behavior of directed percolation with disorder \cite{cardyDirectedPercolationReggeon1980}, and whose Hamiltonian is similar in structure to that of the quantum Ising model.

In practice, the SDRG procedure relies on the disordered nature of a system to iteratively decimate the sites, gaining information about its collective behavior in the process.
In the disordered CP, the SDRG consists of sequential applications of one of the following two steps depending on the momentarily largest parameter in the model.
When the maximal rate is an infection rate between two neighboring sites, $\lambda_{ij}$, the two sites are merged into a ``super site'', as they form a highly correlated cluster.
The effective healing rate of the new super site, $\tilde\mu$, is obtained through a perturbative calculation, since all other rates are smaller:
\begin{subequations}\label{eq:sdrg}
\be
\ln \tilde \mu = \ln \mu_i + \ln\mu_j - \ln\lambda_{ij} + \ln 2\;.
\label{mu_rule}
\ee
In the case when the maximal rate is a healing rate, $\mu_i$, the corresponding site is mostly inactive and can be removed, as long as it is taken into account that weak infection rates are generated perturbatively between its neighbors as
\be
\ln \tilde \lambda_{jk} = \ln \lambda_{ij} + \ln\lambda_{ik} - \ln\mu_i\;.
\label{lambda_rule}
\ee
\end{subequations}
When the SDRG procedure is implemented numerically, the number of sites is reduced by one in each step.
The SDRG procedure is carried out until a single effective site remains, the properties of which are then related to the long-time effective dynamics of the system.

At an IDFP, the logarithmic rates increase in magnitude without limits during the renormalization procedure, therefore the term $\ln 2$ in Eq.~(\ref{mu_rule}) will not influence the asymptotic properties and can be safely omitted \cite{PhysRevLett.76.3001}.
Since the SDRG equations of the disordered quantum Ising model are identical to Eq.~(\ref{eq:sdrg}) without the $\ln 2$ term~\cite{Fisher_1992}, the long-time behavior of the CP is expected to be governed by the same IDFP that controls the critical behavior of the disordered quantum Ising model.
The mapping between the critical points of the two models is conjectured to be asymptotically exact below four dimensions, for large enough system sizes and strong enough initial disorder~\cite{Hooyberghs2003}.

\begin{table*}[t!]
\centering
\caption{\justifying\textbf{Multicritical exponents.} Comparison of the multicritical exponents from (i) Monte Carlo simulations of the multicritical CP (MCP), (ii) quantum Monte Carlo study of the quantum Ising model (QMC), and (iii) strong disorder renormalization group results for the quantum Ising model (SDRG). The exponents are organized as follows, based on how the new estimates presented in this work were obtained: the top four exponents are measured directly through power-law scaling, the next three exponents are directly measured but are impacted by a non-universal microscopic time scale, $t_0$, while the rest are calculated from the values above using the hyperscaling relations listed in Appendix~\ref{sec:scaling}. The numbers in parentheses give an estimate of the uncertainty on the last digits. ``$-$'' is used for exponents of unknown value.}
\label{tab2}
\begin{ruledtabular}
\begin{tabular}{cccccccc}

    & \multicolumn{4}{c}{$d=2$} & & \multicolumn{2}{c}{$d=3$} \\ 
    & MCP \cite{Dahmen2007} & QMC \cite{kramer2025}  & SDRG \cite{Kovacs2022-qmcp}& MCP (this~work) & & SDRG \cite{Kovacs2022-qmcp}& MCP (this~work) \\
    
\noalign{\vskip 1pt}
\hline
\noalign{\vskip 3pt}
         
    $(\psi\bar{\delta})^{-1}$   & 1.08(14) & 1.50(1)  & 1.258(5)   & 1.30(8)  & &0.685(9) & 0.75(8) \\
    $\bar{\Theta}/\bar{\delta}$ & 0.15(28) &  0.99(2) & 0.516(9)  & 0.51(10)  & & 0.05(3) & 0.15(10) \\
    $\psi\bar{\Theta}$          & 0.14(24) & 0.66(1) & 0.410(6) & 0.39(5)  & & 0.08(4) & 0.18(11) \\
    $\beta$                     & 0.81(7)&  0.865(9)  & 1.099(7)  & 1.10(5)  & &1.64(4) & 1.62(20)     \\

\noalign{\vskip 1pt}
\hline
\noalign{\vskip 3pt}
         
    $\bar{\Theta}$              & 0.25(45)& $-$ & 0.579(25) & 0.58(7)    & &0.09(4) & 0.22(14)  \\
    $\psi$                      & 0.57(4) & $-$ & 0.708(20) & 0.698(8)    & &0.93(2) & 0.86(4)  \\
    $\bar{\delta}$              & 1.63(10)& $-$ & 1.12(4) & 1.11(4)    & &1.57(6) & 1.56(9)  \\

\noalign{\vskip 1pt}
\hline
\noalign{\vskip 3pt}
         
    $d_f$                       & 1.07(12)  & 1.332(4)  & 1.205(3)  &  1.23(5)  & & 1.54(2)     & 1.65(14)  \\
    $\nu_\Delta$                & 0.88(10)  & 1.295(5)  & 1.382(7)  & 1.43(15)  & & 1.123(10)   & 1.2(3)   \\
    $\bar{\nu}_\parallel$       & 0.50(9)   & $-$       & 0.98(3)   & 1.00(8)   & & 1.04(3)     & 1.05(19)  \\
     $\nu_p$                    & $-$       & $-$       & 1.168(10) & $-$       & & 0.86(1)     & $-$  \\
    $\gamma$                    & 0.34(14)   & 0.86(3)   & 0.567(11)         & 0.57(14)    & & 0.09(5)    & 0.26(19)  \\

\end{tabular}
\end{ruledtabular}
\end{table*}

Numerically, the one-dimensional disordered CP has been shown to follow SDRG predictions \cite{PhysRevE.72.036126}.
In both two and three dimensions, simulations of the disordered CP at the generic transition \cite{Moreira1996, Vojta2009, Vojta2012} indicate an IDFP with exponents that are close to those obtained by numerical implementations of the SDRG \cite{Kovacs2009-rtfim, Kovacs2010-2d-sdrg, Kovacs2011-sdrg, Kovacs2012-rtfim, Kovacs2011-3d-sdrg} (Table~\ref{tab1}).
However, it is an open problem whether the IDFP is attractive for arbitrarily weak disorder \cite{PhysRevE.78.032101}, or if instead there exists a weak disorder regime, characterized by a line of disorder-dependent fixed points \cite{Hooyberghs2003, Hooyberghs2004, PhysRevE.74.040101, PhysRevE.79.042105}.
So far, the relation between the disordered CP and the quantum Ising model at the multicritical point has not been explored.

In this work, we study the emergent behavior of the disordered CP at the multicritical point in two and three dimensions.
Since the disordered CP is conjectured to be in the same universality class as the disordered quantum Ising model, recent implementations of the SDRG method to the multicritical point of the quantum Ising model \cite{Kovacs2022-qmcp, Kovacs2024-qmcp, Love} are a natural point of comparison for our simulations.
So far, only one study has attempted to characterize the multicritical behavior of the $d=2$ disordered CP through direct simulations \cite{Dahmen2007}.
The results of this investigation are consistent with the activated scaling of an IDFP, but the resulting estimates of the critical exponents are inconsistent with SDRG predictions.
Moreover, the exponents obtained in Ref.~\cite{Dahmen2007}, reported in Table~\ref{tab2}, violate the Harris criterion \cite{PhysRevLett.57.2999}, casting doubts on their accuracy.
While systems that violate the Harris criterion are known to exist \cite{PhysRevLett.113.120602, PhysRevLett.121.100601}, careful testing would be needed to ascertain whether this is the case for the multicritical CP.

Another natural point of comparison for our results are studies of the disordered quantum Ising model that do not rely on the SDRG method.
However, alternative approaches---such as quantum Monte Carlo or tensor network methods---remain extremely challenging and are restricted to relatively small system sizes.
A recent quantum Monte Carlo study of the diluted quantum Ising model \cite{kramer2025} considered $d=2$ systems up to linear size $L=64$, obtaining estimates for a subset of the critical exponents at the multicritical point, reported in Table~\ref{tab2}. However, as acknowledged by the authors, one must be careful when interpreting these results, which are subject to biases that go beyond the reported statistical error.

To the best of our knowledge, no simulations of the $d=3$ multicritical CP have been reported in the literature, and SDRG predictions \cite{Kovacs2022-qmcp, Kovacs2024-qmcp} are the only available results for the three-dimensional disordered quantum Ising model.

Monte Carlo simulations of the CP \cite{Dahmen2007} and quantum Monte Carlo \cite{kramer2025} and SDRG \cite{Kovacs2022-qmcp} studies of the quantum Ising model constitute three independent attempts to characterize the same universality class.
As evident from Table~\ref{tab2}, the resulting exponents at the  $d=2$ multicritical point are typically not consistent within the reported errors.
To address this discrepancy, we present large-scale Monte Carlo simulations of the CP near the multicritical point in two and three dimensions.
The ultra-slow dynamics of an IDFP and strong finite-time effects make resolving the asymptotic behavior in such simulations exceptionally difficult.
Thus, the primary objective of this paper is to ascertain which sets of exponents, if any, are consistent with our Monte Carlo results.

The rest of this paper is organized as follows. In Sec.~\ref{sec:model}, we define the contact process, summarize the relevant scaling expectations, and describe the Monte Carlo simulations. In Sec.~\ref{sec:results}, we present our Monte Carlo results and the extrapolated multicritical exponents in two and three dimensions, followed by a Discussion. Detailed scaling expectations at an IDFP are summarized in Appendix~\ref{sec:scaling}.

\section{Model and Methods}
\label{sec:model}

Let us consider a $d$-dimensional hypercubic lattice of sites, each of which is in one of two states: active (infected) or inactive (healthy).
Time evolution is modeled as a continuous-time Markov chain, where every infected node heals at a rate $\mu$ and every healthy node is infected at a rate $\lambda n/2d$, where $n$ is the number of infected nearest-neighbors.
Without loss of generality, the healing rate $\mu$ can be set to one, leaving the infection rate $\lambda$ as the only parameter controlling the behavior.
For small infection rates, the system is unable to sustain the presence of active sites indefinitely, as the healing rate eventually pushes the dynamics toward a fully inactive state, from which it cannot escape.
On the other hand, when $\lambda$ is large, an infinite system is able to maintain a stationary density of active sites.
The two regimes are separated by a non-equilibrium phase transition that occurs at a critical infection rate $\lambda_c$~\cite{MarroDickman1999}.

When studying the CP, the initial conditions matter, even informing which quantities can be conveniently studied. 
One natural option---used, \textit{e.g.}, in Ref.~\cite{Dahmen2007}---is a fully active initial state.
In this case, the fraction of active sites $\rho(t)$ will start at one and decrease over time, eventually settling at some finite value when $\lambda>\lambda_c$ or decaying to zero when $\lambda<\lambda_c$ \cite{henkel2008non}.
The stationary fraction of active sites plays therefore the role of an order parameter, scaling as a function of the control parameter $\lambda$ according to
\be\label{beta}
\rho(\lambda)\sim|\lambda-\lambda_c|^{\beta}
\ee
for $\lambda\geq\lambda_c$ and close to the critical point.

At the critical infection rate, $\rho(t)$ is typically expected to decay over time as a power-law.
However, in the presence of strong enough quenched disorder, such as dilution of the lattice, this behavior is replaced by the so-called \emph{activated scaling} of an IDFP \cite{IgloiMonthus2005}, in which the fraction of infected sites decays as a power-law of the \emph{logarithm} of time:
\be
    \rho(t)\sim[\ln(t/t_0)]^{-\bar\delta} \;,
\ee
where $t_0$ is a microscopic time scale. For a summary of the relevant scaling expectations at an IDFP, see Appendix~\ref{sec:scaling}.

While fully active initial conditions are common, they are often not the most informative choice.
Instead, in our simulations, we use an initial state containing a single infected site.
In this ``spreading'' case, above the critical infection rate, there is a finite probability that the active cluster will spread from the initial active site to span the entire system. When $\lambda<\lambda_c$, the system will always eventually reach an inactive state.
Note that, due to the stochastic nature of the CP, the system can become trapped in a fully inactive state even when the infection rate is above $\lambda_c$.
In fact, when the infection rate is only slightly larger than $\lambda_c$, most realizations will meet this fate, but a finite fraction, however small, will sustain indefinitely.

In the spreading case, there are three quantities of interest: the average number of active sites $N(t)$, the root-mean-square radius of the active cluster $R(t)$, and the probability $P(t)$ that the system contains at least one active site after time $t$.
The number of active sites is defined as $N=\langle n_a\rangle$, where $n_a$ is the number of active sites in a realization and $\langle\cdot\rangle$ denotes an average over different realizations; the root-mean-square radius is defined as $R=\left[\frac{1}{N}\left\langle\sum_ir^2_i\right\rangle\right]^{1/2}$, where $r_i$ is the distance of the $i$-th active site in a given realization from the initial active site.
Note that alternative definitions of $R$ are sometimes used in the literature, in which the mean-square radius is measured independently in each realization and subsequently averaged~\cite{Juhasz2012}.
We also tested this alternative definition and found the asymptotic scaling to be unchanged, in line with the expectation that there is only one relevant length scale at criticality.

When $\lambda\leq\lambda_c$, $N$, $R$, and $P$ all eventually decay to zero; when $\lambda>\lambda_c$, $N$ and $R$ grow to infinity while $P$ asymptotically approaches a finite value.
At an IDFP, we once again expect the relevant quantities to scale asymptotically as power-laws of the logarithm of time (see Appendix~\ref{sec:scaling} for more details):
\begin{subequations}
\label{scaling}
\begin{align}
    N(t) &\sim [\ln(t/t_0)]^{\bar\Theta}\;,\\
    R(t) &\sim [\ln(t/t_0)]^{1/\psi}\;,\\
    P(t) &\sim [\ln(t/t_0)]^{-\bar\delta}\;.
\end{align}
\end{subequations}

A well-known duality can be established between the spreading and fully active initial conditions \cite{schutz,HooyberghsVanderzande2001}, from which it follows that they share the same critical infection rate $\lambda_c$ and that $\rho(t)$ and $P(t)$ share the same critical exponent $\bar\delta$.

By expressing these observables as functions of each other, we can cancel the dependence on the non-universal microscopic time scale $t_0$, obtaining the following power-law scaling forms:
\begin{subequations}
\label{eq:NPR}
\begin{align}
    R &\sim P^{-1/(\psi\bar\delta)}\;, \\
    N &\sim P^{-\bar\Theta/\bar\delta}\;, \\
    N &\sim R^{\psi\bar\Theta}\;.
\end{align}
\end{subequations}
Note that these exponent combinations are not independent, as they are all related to $\beta/\nu_\Delta$ via (Appendix~\ref{sec:scaling}): 
\begin{subequations}
\label{eq:hyperscaling}
\begin{align}
    \frac{1}{\psi\bar\delta} &= \frac{\nu_\Delta}{\beta} \;, \\
   \frac{\bar\Theta}{\bar\delta} &= d\frac{\nu_\Delta}{\beta}-2 \;, \\
    \psi\bar\Theta &= d-2\frac{\beta}{\nu_\Delta} \;.
\end{align}
\end{subequations}
In turn, Eq.~(\ref{eq:hyperscaling}) implies the hyperscaling formula~\cite{grassbergerLogarithmicCorrections$4+1$dimensional2009}:
\be
\label{eq:hyperscaling_var}
    P^2 R^d / N \sim \mathrm{const} \;.
\ee

In this work, we perform Monte Carlo simulations of the two and three-dimensional CP on a critically diluted lattice according to the following, well-established procedure \cite{Vojta2009,Vojta2012}.
At time $t=0$, the system is initialized with a single active site, then, time evolution is simulated through repeated random updates.
During each update, one active site, $i$, is randomly selected and heals (becomes inactive) with probability $p_\mathrm{heal}=\frac{1}{1+\lambda}$.
If site $i$ does not heal, one of its neighbors, $j$, is randomly selected.
If the bond connecting $i$ and $j$ has not been checked during the simulation prior to this point, it is randomly removed with probability $p_c=\frac{1}{2}$ ($p_c=0.751188$), i.e.,~the bond percolation threshold on a square (cubic) lattice \cite{stauffer_book,PhysRevE.87.052107}.
Then, if site $j$ is inactive and the bond connecting $i$ and $j$ has not been removed, site $j$ is infected (becomes active).
Each update corresponds to a time increment $\Delta t=\frac{1}{n_a}$, where $n_a$ is the current number of active sites in the sample.
This procedure allows us to generate the required bond dilution only in the portion of the lattice that is directly involved in the simulation, speeding up the computation considerably, especially in $d=3$.
The simulations are carried out on large but finite lattices: $2048^2$ sites in $d=2$ and $512^3$ sites in $d=3$.
We check that the active cluster never reaches the boundary of the lattice, so finite-size effects play no role in our results.

We also note that it is common in simulations of the clean CP to use a ``reweighing'' technique to probe several nearby values of the infection rate at the same time~\cite{Dickman1999}.
However, this approach proved unsuccessful when applied to the disordered CP, due to the much longer simulation times~\cite{Vojta2012}.
Therefore, we do not use reweighing in our simulations, and each run is restricted to a single value of $\lambda$.

\section{Results}
\label{sec:results}

\subsection{Critical scaling in two dimensions}\label{sec:2d-critical}

We run at least $1.5\times10^6$ independent simulations for a set of infection rates $3.545\leq\lambda\leq3.5725$ up to a final simulation time $T=10^6$, each using an independently generated critically diluted lattice.
We increase the number of simulations close to the critical region, using at least $7\times10^6$ simulations for near-critical curves.
The number of active sites $N$, the radius of the active cluster $R$, and the fraction of surviving realizations $P$ are measured from the simulations. 

Analyzing these observables directly as functions of time is difficult.
The value of the microscopic time scale $t_0$, which is a priori unknown, can impact any conclusions drawn from finite-time simulations.
Even locating the critical infection rate can pose a challenge. While at criticality the observables are expected to exhibit power-law behavior \emph{asymptotically}, in practice, one finds that finite-time effects make identifying such behavior highly non-trivial.

\begin{figure}[t]   
    \centering
    \includegraphics[width=\linewidth]{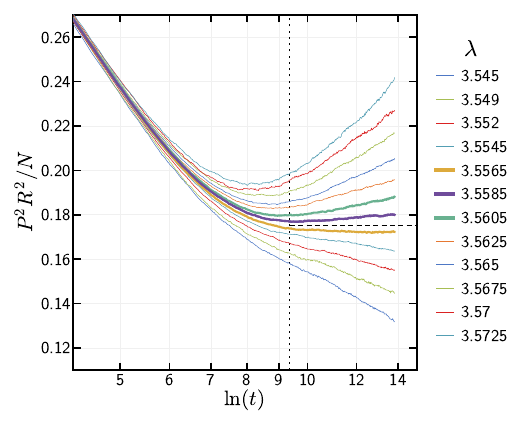}

    \vskip-5mm
    \caption{\justifying
    \textbf{Hyperscaling in $d=2$.}
    We test for hyperscaling using Eq.~(\ref{eq:hyperscaling_var}). Curves in the critical region should appear flat in the limit $t\to\infty$. Thicker lines correspond to the estimated multicritical region around $\lambda_c\approx3.5585$. The vertical dotted line marks the approximate start of the asymptotic regime; the horizontal dashed line is a guide to the eye.
    }
    \label{fig:2d-hyperscaling}
\end{figure}

To help address this problem, we begin by studying the observables combined as in Eq.~(\ref{eq:hyperscaling_var}), which is independent of $t_0$.
The resulting curves are shown in Fig.~\ref{fig:2d-hyperscaling}.
Equation~(\ref{eq:hyperscaling_var}) provides a clear prescription for locating the critical infection rate, since, to be consistent with hyperscaling, only the critical curve should remain constant at long times.
Inspecting Fig.~\ref{fig:2d-hyperscaling}, only the curves corresponding to $\lambda=3.5565$ and $3.5585$ appear approximately flat at long times, which indicates that the tricritical infection rate is in the range $3.5565\leq\lambda_c\leq3.3385$.
Additionally, this analysis allows us to estimate the extent of the asymptotic regime, whose start is marked by a dotted vertical line in Fig.~\ref{fig:2d-hyperscaling}.

\begin{figure*}[t]
    \centering
    \includegraphics[width=\linewidth]{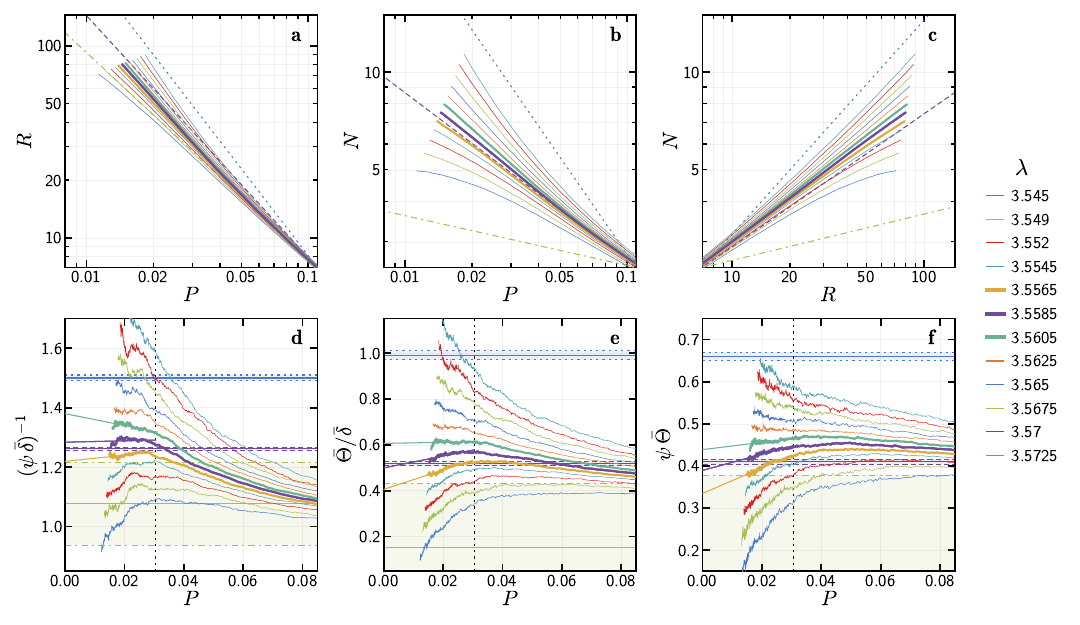}
    
    \begin{subcaptiongroup}
    \phantomcaption\label{fig:2d-RvsP}
    \phantomcaption\label{fig:2d-NvsP}
    \phantomcaption\label{fig:2d-NvsR}
    \phantomcaption\label{fig:2d-RPexp}
    \phantomcaption\label{fig:2d-NPexp}
    \phantomcaption\label{fig:2d-NRexp}
    \end{subcaptiongroup}

    \vskip-5mm
    \caption{\justifying\textbf{Power-law scaling in the $d=2$ model.} The main CP observables are the number of active sites $N$, the root-mean-square radius $R$, and the survival probability $P$.
        \textbf{(a--c)} The observables are plotted as functions of each other to mitigate the effect of the microscopic time scale. 
        Thicker lines correspond to the estimated multicritical region around $\lambda_c\approx3.5585$.
        \textbf{(d--f)} From the slope of these curves, we extract estimates for three exponent combinations: $(\psi\bar\delta)^{-1}$, $\bar\Theta/\bar\delta$, and $\psi\bar\Theta$, see Eq.~(\ref{eq:NPR}). Previous numerical estimates are marked by the purple dashed lines (SDRG predictions \cite{Kovacs2022-qmcp}), green dot-dashed lines (CP Monte Carlo \cite{Dahmen2007}), and blue dotted lines (quantum Monte Carlo \cite{kramer2025}). The vertical dotted lines mark the approximate start of the asymptotic regime.
    }
    \label{fig:2d-NPR}
\end{figure*}

It is also useful when studying the disordered CP to plot $N$, $R$, and $P$ against each other \cite{Vojta2009, Vojta2012}.
Since the same $t_0$ is expected to apply to all three observables, its value becomes irrelevant, allowing for an easier identification of $\lambda_c$.
The resulting plots are shown in Fig.~\ref{fig:2d-RvsP}--\subref{fig:2d-NvsR}.
Each line in these plots traces out the average trajectory of an ensemble of simulations through the space of observables $N$, $R$, and $P$.
Note that $P(t)$ is a monotonically decreasing function of time (the fraction of surviving simulations), so time flows in the direction of decreasing $P$ along each line.
As expected, we observe trajectories diverging from each other over time: simulations below the multicritical infection rate $\lambda_c$ die out ($N,R,P\to0$), while a finite fraction of simulations above $\lambda_c$ approach a stationary density of active sites ($N,R\to\infty$, $P\to\mathrm{const.}$).

To identify the multicritical infection rate, we study the slope of each line as a function of $P$.
Near $\lambda_c$, we should find lines whose slope remains approximately constant as $P\to0$ (i.e., $t\to\infty$ for $\lambda\leq\lambda_c$).
In this limit, the slope of each line corresponds to a combination of critical exponents, as given in Eq.~(\ref{eq:NPR}), 
whose value should be related to the other two exponent combinations as in Eq.~(\ref{eq:hyperscaling}).
To compute the slopes numerically, data for each observable is collected at discrete times during the simulations, defined by $\ln(t_m) = m\ln(1.025)$, with $m=1,2,3,\dots$
Given two observables $\mathcal{O}_1(t)$ and $\mathcal{O}_2(t)$ and an asymptotic scaling relation $\mathcal{O}_2\sim{\mathcal{O}_1}^\zeta$ as $t\to\infty$, we measure an effective exponent at finite times $\zeta_\mathrm{eff}(t)$ as a two-point fit:
\be\label{2pt_fit}
    \zeta_{\mathrm{eff}}(t_m) = \frac{\ln[\mathcal{O}_2(t_{m+\Delta m})] - \ln[\mathcal{O}_2(t_m)]}{\ln[\mathcal{O}_1(t_{m+\Delta m})] - \ln[\mathcal{O}_1(t_m)]}\;,
\ee
where $\Delta m=50$ is chosen to be large enough that the result is not dominated by noise.
The resulting slopes are shown in Fig.~\ref{fig:2d-RPexp}--\subref{fig:2d-NRexp}.

Studying $R$ versus $P$, $N$ versus $P$, and $N$ versus $R$ curves in this manner yields three estimates of the location of $\lambda_c$.
For each pair of observables, we look for curves whose slopes appear approximately flat in the asymptotic regime, as identified in Fig.~\ref{fig:2d-hyperscaling} and marked by a dotted vertical line in Fig.~\ref{fig:2d-RPexp}--\subref{fig:2d-NRexp}.
Analyzing the pairwise scaling of the observables in this manner, we identify three likely ranges for the location of the critical infection rate: $R$ vs.~$P$ curves are consistent with $3.5565\leq\lambda_c\leq3.5625$ (Fig.~\ref{fig:2d-RPexp}), $N$ vs.~$P$ curves are consistent with the same range of $3.5565\leq\lambda_c\leq3.5625$ (Fig.~\ref{fig:2d-NPexp}), while the $N$ vs.~$R$ curves are consistent with $3.5585\leq\lambda_c\leq3.565$ (Fig.~\ref{fig:2d-NRexp}).

\begin{figure*}[t]
    \centering
    \includegraphics[width=\linewidth]{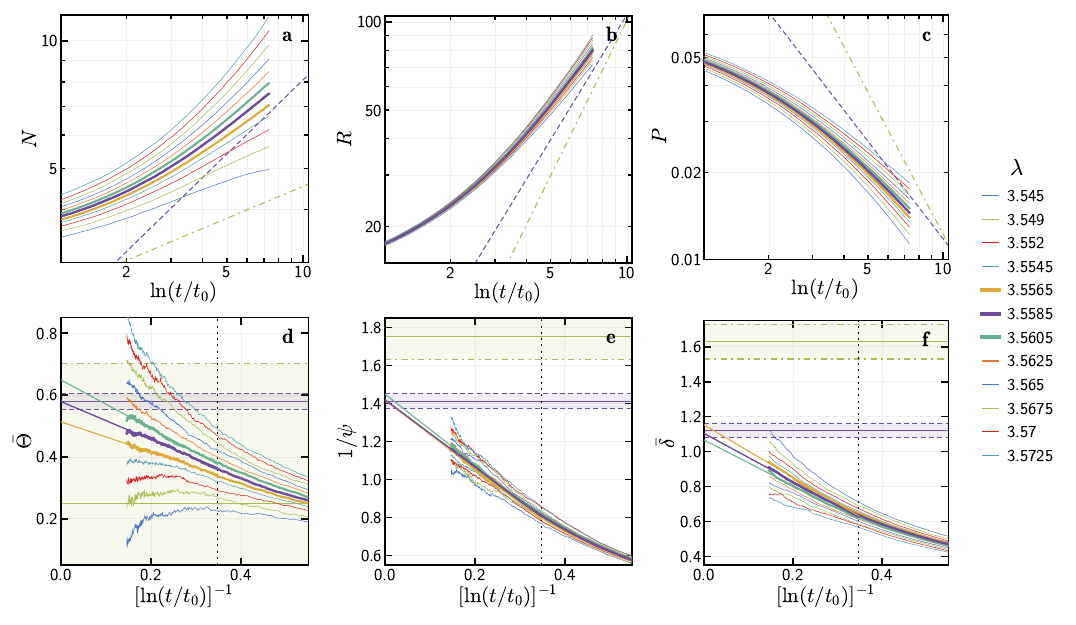}
    
    \begin{subcaptiongroup}
    \phantomcaption\label{fig:2d-Nvst}
    \phantomcaption\label{fig:2d-Rvst}
    \phantomcaption\label{fig:2d-Pvst}
    \phantomcaption\label{fig:2d-Nexp}
    \phantomcaption\label{fig:2d-Rexp}
    \phantomcaption\label{fig:2d-Pexp}
    \end{subcaptiongroup}

    \vskip-5mm
    \caption{\justifying\textbf{Dynamical scaling in $d=2$.}
        Each observable is plotted against $\ln(t/t_0)$ using an estimate of $\ln(t_0)=6.5$.
        All three exponents match the SDRG predictions within the uncertainty. The purple dashed lines mark the SDRG prediction \cite{Kovacs2022-qmcp}; the green dot-dashed lines mark the CP Monte Carlo results of Ref.~\cite{Dahmen2007}. The vertical dotted lines mark the approximate start of the asymptotic regime.
        }
    \label{fig:2d-NPR-t0}
\end{figure*}

To summarize, hyperscaling (Fig.~\ref{fig:2d-hyperscaling}) and pairwise scaling of the observables (Fig.~\ref{fig:2d-NPR}) lead to slightly different but overlapping ranges of estimates for $\lambda_c$, with $\lambda=3.5585$ consistently appearing as a good candidate critical curve across all figures.
Therefore, we estimate the location of the multicritical infection rate of the two-dimensional CP to be $\lambda_{c}^{(2)}=3.5585(20)$.

Based on this estimate of $\lambda_c$, we carry out $P\to0$ extrapolations of the effective exponents obtained from pairwise scaling of the observables in the proposed critical region to estimate $(\psi\bar\delta)^{-1}$, $\bar\Theta/\bar\delta$, and $\psi\bar\Theta$ (Fig.~\ref{fig:2d-RPexp}--\subref{fig:2d-NRexp}).
The extrapolations are done using $4\Delta m=200$ data points for each curve, corresponding to the approximate extent of the asymptotic regime.
The numerical results for the three exponent combinations can be found in Table~\ref{tab2}, showing good agreement with the SDRG predictions.

We turn now to the task of estimating the microscopic time scale $t_0$ and measuring the critical exponents $\bar\Theta$, $\psi$, and $\bar\delta$ individually.
We do not have a direct method to estimate $t_0$.
However, we can rely on our estimate of the critical infection rate.
Based on the observation that pairwise scaling of the observables is consistent with SDRG predictions, we consider whether there exists a unique value of $t_0$ for which all three critical exponents, $\bar\Theta$, $\psi$, and $\bar\delta$ are in turn consistent with SDRG predictions.
We address this question by studying the slopes of $N$, $R$, and $P$ as functions of $\ln(t/t_0)$ in the limit $t\to\infty$ (Fig.~\ref{fig:2d-NPR-t0}).
To study the asymptotic scaling, we calculate effective exponents using Eq.~(\ref{2pt_fit}) and extrapolate the result to the limit $[\ln(t/t_0)]^{-1}\to0$.

We find that curves in the proposed critical region consistently yield exponent values in line with SDRG predictions using $\ln(t_0)=6.5$, which is similar to previous estimates of the microscopic time scale in the bond-diluted CP \cite{Vojta2009}.
Figure~\ref{fig:2d-NPR-t0} shows plots of $N(t)$, $R(t)$, and $P(t)$ and their respective slopes using $\ln(t_0)=6.5$.
The resulting numerical estimates for $\bar\Theta$, $\psi$, and $\bar\delta$ are reported in Table~\ref{tab2}.
Note that, as the value of $t_0$ was informed by 
the SDRG predictions, the resulting critical exponents reported in Table~\ref{tab2} are subject to systematic uncertainty that exceeds the reported statistical error.

\subsection{Critical scaling in three dimensions}\label{sec:3d-critical}

\begin{figure}[t]    
    \centering
    \includegraphics[width=\linewidth]{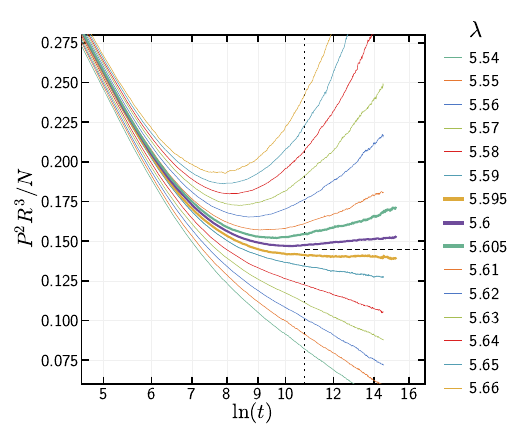}

    \vskip-5mm
    \caption{
    \label{fig:3d-hyperscaling}
    \justifying
    \textbf{Hyperscaling in $d=3$.}
    We test for hyperscaling using Eq.~(\ref{eq:hyperscaling_var}).
    Curves in the critical region should appear flat in the limit $t\to\infty$. 
    Thicker lines correspond to the estimated multicritical region around $\lambda_c\approx5.6$.
    The vertical dotted line marks the approximate start of the asymptotic regime; the horizontal dashed line is a guide to the eye.
    }
\end{figure}

\begin{figure*}[ht]
    \centering
    \includegraphics[width=\linewidth]{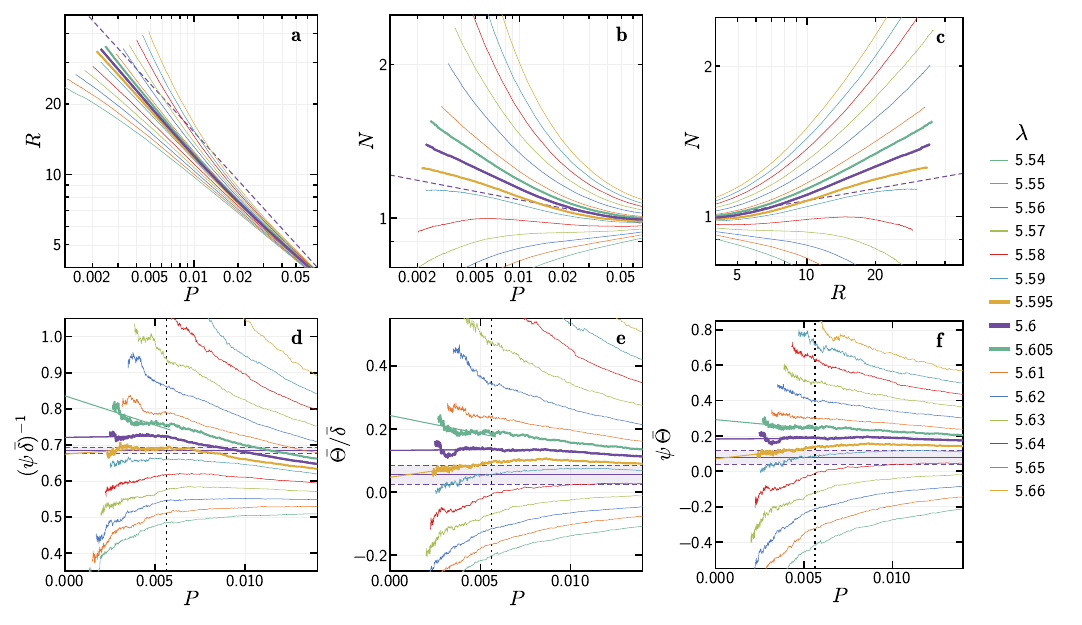}
    
    \begin{subcaptiongroup}
    \phantomcaption\label{fig:3d-RvsP}
    \phantomcaption\label{fig:3d-NvsP}
    \phantomcaption\label{fig:3d-NvsR}
    \phantomcaption\label{fig:3d-RPexp}
    \phantomcaption\label{fig:3d-NPexp}
    \phantomcaption\label{fig:3d-NRexp}
    \end{subcaptiongroup}
    
    \vskip-5mm
    \caption{\justifying\textbf{Power-law scaling in the $d=3$ model.} The standard observables are plotted as functions of each other.  
    Thicker lines correspond to the estimated multicritical region around $\lambda_c\approx5.6$.
    From the slope of these curves, we extract estimates for the exponent combinations $(\psi\bar\delta)^{-1}$, $\bar\Theta/\bar\delta$, and $\psi\bar\Theta$. The dashed purple lines mark the SDRG prediction \cite{Kovacs2022-qmcp}.
    The vertical dotted lines mark the approximate start of the asymptotic regime.
    }
    \label{fig:3d-NPR}
\end{figure*}

To study the critical behavior of the three-dimensional multicritical CP, we run at least $4\times10^6$ independent simulations for a set of infection rates $5.54\leq\lambda\leq5.66$, up to a final simulation time of $T = 2\times10^6$.
We increase the final time to $T = 4\times10^6$ and the number of simulations to at least $4\times10^7$ for lines close to our estimate of the multicritical point.
The analysis of the three-dimensional simulations is analogous to the two-dimensional case, as detailed in Section \ref{sec:2d-critical}.

We begin by studying hyperscaling as prescribed in Eq.~(\ref{eq:hyperscaling_var}).
Fig.~\ref{fig:3d-hyperscaling} indicates that the curves corresponding to $\lambda=5.595$ and $5.6$ appear approximately flat at long times, allowing us to estimate the extent of the asymptotic regime, marked by the vertical dotted line.
Next, we study the pairwise scaling of the observables as a function of $P$ (Fig.~\ref{fig:3d-NPR}).
As in $d=2$, we look for curves with slopes that remain approximately constant in the asymptotic regime, whose approximate extent, as identified from Fig.~\ref{fig:3d-hyperscaling}, is marked by the vertical dotted line in Fig.~\ref{fig:3d-RPexp}--\subref{fig:3d-NRexp}.
We identify in this way three likely ranges for the multicritical infection rate: $R$ vs.~$P$ curves are consistent with $5.58\leq\lambda_c\leq5.61$ (Fig.~\ref{fig:3d-RPexp}), $N$ vs.~$P$ curves are consistent with $5.595\leq\lambda_c\leq5.61$ (Fig.~\ref{fig:3d-NPexp}), and $N$ vs.~$R$ curves are consistent with the same range of $5.595\leq\lambda_c\leq5.61$ (Fig.~\ref{fig:3d-NRexp}).
Taking into account the estimated ranges for the critical infection rate provided by hyperscaling (Fig.~\ref{fig:3d-hyperscaling}) and pairwise scaling of the observables (Fig.~\ref{fig:3d-NPR}), we estimate the multicritical infection rate of the three-dimensional disordered CP to be $\lambda_c^{(3)}=5.600(5)$.

Based on this estimate of $\lambda_c$, we carry out $P\to0$ extrapolations of the pairwise scaling exponents of near-critical curves in the asymptotic regime to estimate $(\psi\bar\delta)^{-1}$, $\bar\Theta/\bar\delta$, and $\psi\bar\Theta$ (Fig.~\ref{fig:3d-RPexp}--\subref{fig:3d-NRexp}).
The resulting estimates of the three exponent combinations can be found in Table~\ref{tab2} and all agree with SDRG predictions within the uncertainty.

\begin{figure*}[t]
    \centering
    \includegraphics[width=\linewidth]{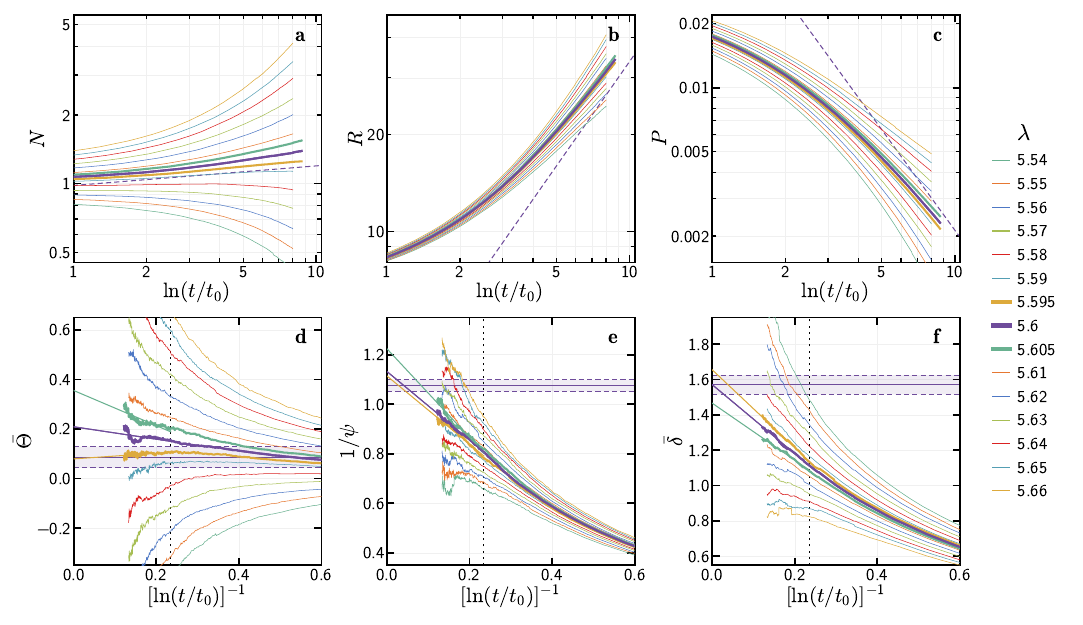}
    
    \begin{subcaptiongroup}
    \phantomcaption\label{fig:3d-Nvst}
    \phantomcaption\label{fig:3d-Rvst}
    \phantomcaption\label{fig:3d-Pvst}
    \phantomcaption\label{fig:3d-Nexp}
    \phantomcaption\label{fig:3d-Rexp}
    \phantomcaption\label{fig:3d-Pexp}
    \end{subcaptiongroup}
    
    \vskip-5mm
    \caption{\justifying\textbf{Dynamical scaling in $d=3$.} 
     Each observable is plotted against $\ln(t/t_0)$ using $\ln(t_0)=6.5$, the same microscopic time scale used in $d=2$.
     The infinite-time ($[\ln(t/t_0)]^{-1} \to 0$) extrapolations are consistent with SDRG results for all three exponents (dashed purple lines). The vertical dotted lines mark the approximate start of the asymptotic regime.
    }
    \label{fig:3d-NPR-t0}
\end{figure*}

Next, we study each observable separately as a function of time.
The value of the microscopic time scale $t_0$ is not expected to depend on dimensionality, as long as the model and type of disorder are kept the same.
Accordingly, we find that the slopes of all three observables, extrapolated to the limit $[\ln(t/t_0)]^{-1}\to0$, are consistent with SDRG results if the time scale is set to $\ln(t_0)=6.5$, the same value used in two dimensions (Fig.~\ref{fig:3d-NPR-t0}).
The resulting estimates of the critical exponents are reported in Table~\ref{tab2}.
Just like in $d=2$, these estimates depend on the value of $t_0$, hence they are subject to systematic uncertainty that exceeds the reported statistical error.

\subsection{Off-critical scaling}
To access more critical exponents, we estimate $\beta$ by studying the scaling of the observables away from the critical infection rate (see Eq.~(\ref{beta})).
We again use a spreading initial condition. 
Using the alternative, fully-active initial state would be more restrictive as the stationary density of active sites is the only accessible observable, scaling with the control parameter $\lambda$ as $\rho(\lambda)\sim\vert\lambda-\lambda_c\vert^\beta$.
Duality implies \cite{schutz,HooyberghsVanderzande2001} that in the case of spreading initial conditions, the survival probability $P(\lambda)$ follows the same scaling.
However, going directly after the dependency of either $\rho$ or $P$ on the infection rate requires measuring the infinite-time stationary behavior close to the critical point, which is made challenging by the ultra-slow dynamical scaling.
To get around this problem, we follow Ref.~\cite{Vojta2012} and study the derivative of the logarithm of the observables with respect to $\lambda$.
The relevant scaling form for this analysis is:
\be\label{eq:beta_scaling}
    \left.\frac{\partial\ln N}{\partial\lambda}\right\vert_{\lambda=\lambda_c} \sim P^{-1/\beta}\;.
\ee
Analogous formulas hold for $\partial_\lambda\ln P$ and $\partial_\lambda\ln R$.
These formulas are especially useful since they do not require knowledge of $t_0$.

\begin{figure*}[t]
    \centering
    \includegraphics[width=\linewidth]{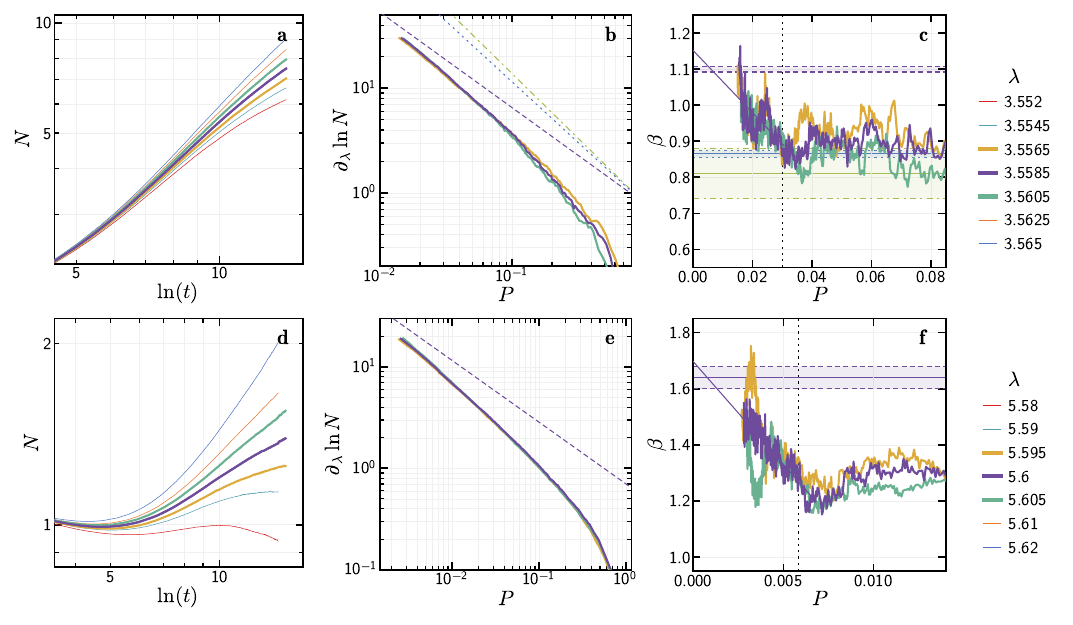}
        
    \begin{subcaptiongroup}
    \phantomcaption\label{fig:beta-2d-Nvst}
    \phantomcaption\label{fig:beta-2d-dNvsP}
    \phantomcaption\label{fig:beta-2d-exp}
    \phantomcaption\label{fig:beta-3d-Nvst}
    \phantomcaption\label{fig:beta-3d-dNvsP}
    \phantomcaption\label{fig:beta-3d-exp}
    \end{subcaptiongroup}
    
    \vskip-5mm
    \caption{\justifying\textbf{Off-critical scaling in $d=2$ and $d=3$.} Estimation of the critical exponent $\beta$ in $d=2$ (a--c) and $d=3$ (d--f). We calculate the rate $\partial_\lambda\ln N$ at which near-critical $N(\lambda,t)$ curves diverge from a candidate critical curve as a function of the infection rate $\lambda$ (panels b and e). $\beta$ is then calculated from the slope of $\partial_\lambda\ln N$ plotted against $P$ (panels c and f).
    Previous numerical estimates are marked by the purple dashed lines (SDRG predictions \cite{Kovacs2022-qmcp}), green dot-dashed lines (CP Monte Carlo \cite{Dahmen2007} in $d=2$), and blue dotted lines (quantum Monte Carlo \cite{kramer2025} in $d=2$).
    The vertical dotted lines in panels c and f mark the approximate start of the asymptotic regime, as determined from Figs.~\ref{fig:2d-hyperscaling} and \ref{fig:3d-hyperscaling}.
    }
    \label{fig:beta}
\end{figure*}

Our analysis of the off-critical scaling behavior proceeds as follows. 
First, we consider a set of near-critical $N(\lambda,t)$ curves and use their distance from a candidate critical curve to estimate $\partial_\lambda\ln N$ as a function of time.
To increase the accuracy of our estimate, we average the result over the four curves closest to the candidate critical curve.
$\partial_\lambda\ln N$ is then plotted against $P$, and the slope of the resulting curve is used to estimate $\beta$ by extrapolating to $P\to0$. The full procedure is summarized in Fig.~\ref{fig:beta}.

In the $d=2$ model, we perform this analysis separately for all three curves within the critical region ($3.5565\leq\lambda\leq3.5605$).
We find that the slope of $\partial_\lambda\ln N$ as a function of $P$ is initially close to the CP Monte Carlo results of Ref.~\cite{Dahmen2007} and to the quantum Monte Carlo results of Ref.~\cite{kramer2025}.
However, as $P\to0$, we consistently observe a crossover to a different scaling behavior (Fig.~\ref{fig:beta-2d-exp}), which occurs when the system reaches the asymptotic regime, as identified in Figs.~\ref{fig:2d-hyperscaling} and \ref{fig:3d-hyperscaling}.
We perform a linear extrapolation of the slope of $\partial_\lambda\ln N$ as $P\to0$ using the same range of data as previous extrapolations, which yields a value of $\beta$ consistent with the SDRG prediction \cite{Kovacs2022-qmcp}.
The resulting numerical estimate of $\beta$ is reported in Table~\ref{tab2}, where the primary source of the reported statistical error is the uncertainty on the value of $\lambda_c^{(2)}$. We note that our multicritical estimate of the order-parameter exponent $\beta$ is compatible with numerical estimates of $\beta$ at the generic transition of the $d=2$ disordered CP \cite{Vojta2009, Kovacs2011-3d-sdrg} (Table~\ref{tab1}).

The same analysis is carried out for each curve in the proposed critical region of the $d=3$ model ($5.595\leq\lambda\leq5.605$), where, similarly to $d=2$, we consistently observe a crossover to a different scaling regime as $P\to0$.
The resulting numerical estimate---reported in Table~\ref{tab2}---is consistent with the multicritical value predicted by the SDRG \cite{Kovacs2022-qmcp}.
Note that carrying out the same analysis using $\partial_\lambda\ln P$ or $\partial_\lambda\ln R$ in place of $\partial_\lambda\ln N$ yields similar, if noisier, results.

\section{Discussion}
\label{sec:discussion}

In this work, we have determined the critical exponents of the multicritical disordered CP through large-scale Monte Carlo simulations in both two and three dimensions. In two dimensions, there have been three independent estimates of the multicritical behavior: the Monte Carlo simulation of Ref.~\cite{Dahmen2007} and two studies of the disordered quantum Ising model (expected to be in the same universality class) using the strong disorder renormalization group (SDRG) method~\cite{Kovacs2022-qmcp} and a quantum Monte Carlo simulation~\cite{kramer2025}, respectively.
In three dimensions, to the best of our knowledge, the only previous estimates of the multicritical exponents were obtained through the SDRG~\cite{Kovacs2022-qmcp}.
The determination of scaling exponents in the multicritical CP is impaired by the ultra-slow activated scaling, large microscopic time scales, and strong finite-time corrections. Similar strong sub-leading corrections have been reported in the four-dimensional clean CP~\cite{janssenLogarithmicCorrectionsDirected2004,grassbergerLogarithmicCorrections$4+1$dimensional2009}---\textit{i.e.}, at the upper critical dimension of the DP universality class.
Here, we presented a methodology that leverages infinite-time extrapolations and hyperscaling to arrive at consistent estimates of the critical exponents.
In both $d=2$ and $d=3$, our exponent estimates are within the error consistent with those obtained by the SDRG method for the multicritical quantum Ising model \cite{Kovacs2022-qmcp}, which serves as the first confirmation of the multicritical SDRG results and indicates that the approximations used in the numerical SDRG scheme are indeed valid. This opens the possibility that our methodology could also show an improved agreement between CP simulations and the SDRG results at the generic transition, see Table~\ref{tab1}.

In two dimensions, our exponents satisfy the Harris criterion, a qualitative difference from those obtained by Dahmen et al.~\cite{Dahmen2007}. 
Note that in Ref.~\cite{Dahmen2007}, the Monte Carlo simulations only used the largest connected component of each critically diluted lattice. 
However, as discussed in Ref.~\cite{Kovacs2022-qmcp}, this methodological difference only leads to a small  shift in a subset of the exponents. 
The main factor explaining the difference between our results and those of Ref.~\cite{Dahmen2007} is their choice of a fully infected initial condition.
In this type of simulation, the density of active sites is the only accessible observable, and any interpretation of the results, including locating the critical region, is significantly impacted by the microscopic time scale.
As for the quantum Monte Carlo results of Ref.~\cite{kramer2025}, future work would ideally improve on three aspects: (i) obtain larger sizes for the estimates to converge, (ii) characterize the remaining exponents---most importantly quantifying $\psi$ to check the presence of activated scaling---and (iii) improve the reported error bars by incorporating systematic effects in addition to the currently included statistical error.

The observation that the $d=2$ and $d=3$ disordered (critical and multicritical) CP and the quantum Ising model likely belong to the same universality class has widespread consequences. Prominently, it opens a direct way to measure equilibrium quantum
effects in the classical out-of-equilibrium CP, as the dynamics is governed by the same underlying activity clusters \cite{kovacs-2020}.

\begin{acknowledgments}
We are grateful to P.~Grassberger for fruitful discussion and for helping to substantially improve our numerical results.
We thank W.~T.~Engedal and R.~Juh\'asz for helpful comments and discussion.
This work was supported by the U.~S.~National Science Foundation under Grant No.~PHY-2310706 of the QIS program in the Division of Physics, and Northwestern's Office of Undergraduate Research with the generous support of the Hung-Farinelli Family under the 2023 Summer Undergraduate Research Grant.
This research was supported in part through the computational resources and staff contributions provided for the Quest high performance computing facility at Northwestern University, which is jointly supported by the Office of the Provost, the Office for Research, and Northwestern University Information Technology.
L.~V.~L.~and I.~A.~K.~acknowledge support from the NSF–Simons National Institute for Theory and Mathematics in Biology, jointly funded by the U.S.~National Science Foundation (Award No.~2235451) and the Simons Foundation (Award No.~MP-TMPS-00005320).
\end{acknowledgments}

\newpage
\bibliography{References}

\appendix
\section{Scaling expectations at an IDFP}
\label{sec:scaling}

Here, we provide a summary of the relevant scaling relations at an IDFP, following Refs.~\cite{Hooyberghs2003, Hooyberghs2004}.
The control parameter is traditionally chosen as $\Delta_0\equiv \overline{\ln(\lambda/\mu)}$ \cite{IgloiMonthus2005}, with the overbar standing for a disorder average. Note that here, we use $\mu=1$.
The deviation from the critical point $\Delta_0^*$ is then given by 
$\Delta\equiv \Delta_0-\Delta_0^*$. 
The activity density $\rho$ of the system can be chosen as the order parameter, vanishing as 
$
\rho(\Delta)\sim\Delta^{\beta} 
$
close to the transition in the active phase.
As usual, the spatial correlation length $\xi_{\Delta}$ diverges close to the critical point as 
$
\xi_{\Delta}\sim |\Delta|^{-\nu_{\Delta}}
$, 
with the correlation length exponent $\nu_{\Delta}$. 
At an IDFP, the time and length scales
are related to each other as 
\be
\ln\xi_{\parallel}\sim\xi_{\Delta}^{\psi},
\label{act} 
\ee
with $\psi$ being the tunneling exponent, leading to $\overline{\nu}_{\parallel}=\nu_\Delta\psi$.
The order parameter for finite size $L$ and time $t$ is expected to transform as 
\be
\rho(L,t,\Delta)=b^{-x}\tilde\rho(L/b,\ln t/b^{\psi},\Delta b^{1/\nu_{\Delta}}),
\label{order}
\ee 
when the length is rescaled by a factor $b$. Hence, we obtain the relation $x\equiv \beta/\nu_{\Delta}$. Then, the fractal dimension is given by $d_f=d-x$.
The survival probability is expected to scale as
the order parameter \cite{henkel2008non,RevModPhys.76.663}
\be 
P(L,t,\Delta)=b^{-x}\tilde P(L/b,\ln t/b^{\psi},\Delta b^{1/\nu_{\Delta}}). 
\label{P_scale} 
\ee  
With the binary variables $n_i$ coding active ($n_i=1$)
and inactive ($n_i=0$) sites, the spatio-temporal correlation function $C[n_0(t=0),n_i(t)]$ summed over the position $i$ yields for the scaling of the average number of active nodes: 
\be  
N(L,t,\Delta)=b^{d-2x}\tilde N(L/b,\ln t/b^{\psi},\Delta b^{1/\nu_{\Delta}}).\label{N_scale} 
\ee 
The radius is defined as the root-mean-square of the
distance of active nodes from the origin, averaged for the surviving samples up to time $t$. 
Using the same scaling form as the spatio-temporal correlation function we obtain 
\be  
R(L,t,\Delta)=b\tilde R(L/b,\ln t/b^{\psi},\Delta b^{1/\nu_{\Delta}}).
\label{R_scale} 
\ee 
At the critical point ($\Delta=0$) of the infinite system ($L=\infty$), it follows from the above relations that the observables depend on time asymptotically as given by Eq.~(\ref{scaling}), 
with the exponents given in terms of the earlier ones as 
\be
\bar\delta\equiv x/\psi
\label{deltax}
\ee 
and the hyperscaling relation 
\be\label{thetax}
\bar\Theta\equiv (d-2x)/\psi\;.
\ee
The hyperscaling formula reported in Eq.~(\ref{eq:hyperscaling_var}) follows directly from Eqs.~(\ref{P_scale}), (\ref{N_scale}), and (\ref{R_scale}).
From Eq.~(\ref{thetax}) and the hyperscaling relation $\gamma=d\nu_\Delta-2\beta$~\cite{goldenfeld_book}, we also have
\be\label{eq:gamma}
    \gamma=\bar\Theta\,\psi\,\nu_\Delta\;.
\ee
At the multicritical IDFP, there are two relevant directions, corresponding to two control parameters. In addition to moving away from the multicritical point by changing the infection rates (controlled by $\Delta$), we can also move away by changing the dilution parameter, $p$, leading to an additional correlation length exponent $\nu_p$:
\be 
\xi_{p-p_c}\sim |p-p_c|^{-\nu_{p}}.
\ee

\end{document}